\documentclass[12pt]{article}

\usepackage{amsmath,amsfonts}

\usepackage{slashed}

\usepackage{cite}

\advance \textheight by \topskip
\makeatletter
\@addtoreset{equation}{section}
 
\makeatother

\newcommand{\be}{\begin{equation}}
\newcommand{\ee}{\end{equation}}
\newcommand{\bea}{\begin{eqnarray}}
\newcommand{\eea}{\end{eqnarray}}
\newcommand{\dd}{\partial}

\def\>{\rangle}
\def\<{\langle}

\begin{document}

\title{
{\bf A note on Weyl gauge symmetry in gravity }}


\author{
{\sf   N. Mohammedi} \thanks{e-mail:
noureddine.mohammedi@univ-tours.fr}$\,\,$${}$
\\
{\small ${}${\it Institut  Denis Poisson (CNRS - UMR 7013),}} \\
{\small {\it Universit\'e  de Tours,}}\\
{\small {\it Facult\'e des Sciences et Techniques,}}\\
{\small {\it Parc de Grandmont, F-37200 Tours, France.}}}
\date{}
\maketitle
\vskip-1.5cm

\vspace{2truecm}

\begin{abstract}

\noindent
A scale invariant theory of gravity, containing at most two derivatives, requires, in addition to the Riemannian metric,  a scalar field 
and (or) a gauge field. The gauge field is usualy used to construct the  affine connection of Weyl geometry. In this note,  we incorporate
both the gauge field and the scalar field to build a generalised scale invariant Weyl affine connection. The Ricci tensor and the Ricci scalar made out of this
generalised Weyl affine connection contain, naturally,  kinetic terms for the scalar field and the gauge field. This provides a geometric interpretation
for these terms. It is also shown that scale invariance in the presence of a cosmological constant and mass terms is not completely 
lost. It becomes a duality transformation relating various fields.

\end{abstract}

\newpage

%

\setcounter{equation}{0}

\section{Introduction}

It is plausible to think that at very high energies, when gravity was strong, there were
only massless fields in the Standard Model of particle physics.
Hence the Standard Model coupled to gravity might have enjoyed a scale symmetry. 
This symmetry would then be of importance to the physics of the early Universe
and its effects could be seen in cosmological observations.
This explains also the renewed research activity in this direction. {}A history and a modern review 
of scale invariant theories of gravity could be found in \cite{rev1} (see also
\cite{rev2}, \cite{rev3}, \cite{rev4}, \cite{rev5}). Some issues regarding the use of 
scale invariant gravity in cosmology, inflation and other astrophysical observations  are treated in 
\cite{ob1,ob2,ob3,ob4,ob5,ob6,ob7,ob8,ob9,ob10,ob11,ob12,ob13,ob14,ob15,ob16,ob17,ob18,ob19,ob20,ob21,ob22,ob23,ob24,ob25,ob26,ob27}.
Various studies concerning the coupling of the Standard Model of particle physics to gravity in a scale invariant
manner are accounted for in \cite{sm1,sm2,sm3,sm4,sm5,sm6,sm7,sm8,sm9,sm10,sm11,sm12,sm13}.
Classical solutions in the context of scale invariant gravity could be found in \cite{c1}, \cite{c2}, \cite{c3}, \cite{sm11}.
\par
As it is well-known, in order to render Einstein theory of gravity invariant under the 
rescaling of the metric tensor 
\be
g_{\mu\nu} \longrightarrow e^{\sigma(x)}\, g_{\mu\nu} \,\,\,\,\,
\ee
one needs the introduction of a scalar fiels $\varphi$ having the transformation 
\be
\varphi \longrightarrow \varphi +\frac{1}{q}\,{\sigma(x)} \ \,\,\,\,\,.
\label{scalat-tran}
\ee
The scale invariant theory of gravity  is then described by the action\footnote{Our  conventions  are such
the Riemann tensor is given by 
$R^\mu_{\nu\rho\sigma}=\dd_\rho \Gamma^\mu_{\nu\sigma} +
\Gamma^\mu_{\rho\alpha}\Gamma^{\alpha}_{\nu\sigma} - \left(
\rho \leftrightarrow \sigma\right)$, the Ricci tensor is   $R_{\mu\nu}
=R^\alpha_{\mu\alpha\nu}$ and the Ricci scalar is $R=g^{\mu\nu}R_{\mu\nu}$. The signature of the  metric $g_{\mu\nu}$ is mostly minus $\left(+\,,\,-\,,\,-\,,\,-\right)$.
The covariant derivative is denoted $\nabla_\mu$ such that $\nabla_\mu V^\nu = \dd_\mu V^\nu +\Gamma^\nu_{\mu\sigma}V^\sigma$, for a four-vector $V^\nu$,
and $\nabla^2=\nabla_\mu \nabla^\mu$. }
\bea
S_1  &=& 
\xi\,\int d^4x\sqrt{-g}\,e^{-q\varphi}\Bigg\{- R - \left( 3q\,\nabla^2\varphi -\frac{3}{2}q^2\,\dd_\mu\varphi \dd^\mu\varphi\right)
+\kappa\,e^{-q\varphi} \Bigg\}
\label{ghost}
\,\,\,\,\,.
\eea
The last term is independently scale invariant and $\xi$, $\kappa$ and $q$ are constants.
\par
The action  (\ref{ghost}) can be cast in the form
\bea
S  &=& 
\xi\,\int d^4x\sqrt{-G}\,\Bigg\{- R\left(G\right)
+\kappa  \Bigg\}
\label{ghost-1}
\,\,\,\,\,,
\eea
where the scale invariant metric $G_{\mu\nu}$ is defined as
\be
G_{\mu\nu} = e^{-q\varphi} \,g_{\mu\nu} \,\,\,\,\,
\ee
and $ R\left(G\right)$ is the Ricci scalar constructed out of this metric. Hence, the field $\varphi$ 
is simply the conformal factor of the metric $G_{\mu\nu}$.
\par 
In the action (\ref{ghost}), the propagating scalar field is in fact
\be
\chi =\exp{\left({-\frac{q}{2}\,\varphi}\right)}\,\,\,\,\,.
\label{chi}
\ee
{}For convenience, we still work with the field $\varphi$ through out this note. The kinetic term
(after integration by parts) for the scalar field $\chi$ is 
\be
-6\xi \sqrt{-g}\, \dd_\mu \chi  \dd^\mu \chi  \,\,\,\,\,.
\label{chi-1}
\ee
As $\xi$ is a positive constant, this kinetic terms is negative and leads to the interpretation that 
the scalar field $\chi$ is a ghost. 
\par
There is in fact another manner to obtain a scale invariant theory of gravity. It consists 
in using  a gauge field $C_\mu$ having the transformation \cite{rev1}
\bea
C_\mu &\longrightarrow&  C_\mu + \frac{1}{{q}'}\,\dd_\mu \sigma(x)\,\,\,\,\,.
\label{}
\eea
Here $q'$ is a constant. The scale invariant action in this case is given by
\bea
S_2  &=& 
\xi\,\int d^4x\sqrt{-g}\,e^{-q\varphi}\Bigg\{- R 
  - \left( 3{q'}\,\nabla_\mu C^\mu -\frac{3}{2}{{q}'}^2\,C_\mu C^\mu \right)+\kappa\,e^{-q\varphi} \Bigg\}
\,\,\,.
\label{Ghilencea-1-1}
\eea
Notice that one still needs the scalar field $\varphi$.  It has the scale transformation (\ref{scalat-tran}).
\par
The two action $S_1$ and $S_2$ are in fact related. The gauge field  $C_\mu$ in (\ref{Ghilencea-1-1}) is a non-propagating
field whose equation of motion is
\be
C_\mu=\frac{q}{q'}\,\dd_\mu\varphi \,\,\,.
\ee
Injecting this in (\ref{Ghilencea-1-1}), we recover the action $S_1$ in  (\ref{ghost}).
\par
The two scale invariant theories described by the actions $S_1$ and $S_2$ have two
drawbacks. The first is that the propagating field $\chi$ in  $S_1$ is a ghost field
having a negative kinetic term as shown in (\ref{chi-1}). The second is that the  gauge
field  $C_\mu$ in $S_2$ does not propagate. Fortunately, the remedy to this two problems is 
known in the literature. It consists in considering the  scale invariant actions \cite{sm2}
\bea
S^{(a)}_{\text{scale}}  &=& S_a 
 + \sigma_a\, \int d^4x\sqrt{-g}\,e^{-q\varphi}\, {\cal D}_\mu \varphi {\cal D}^\mu \varphi
\nonumber \\
&-& \frac{1}{4} \int d^4x\sqrt{-g}\,{C_{\mu\nu}\,C^{\mu\nu}}
\,\,\,\,\,,
\label{Ghilencea-2}
\eea
where 
\bea
C_{\mu\nu} &=& \dd_\mu C_\nu -  \dd_\nu C_\mu \,\,\,\,\,
\eea 
is the fields strenght corresponding to the  gauge field $C_\mu$ and  
\bea
{\cal D}_\mu \varphi &=& \dd_\mu \varphi -\frac{q^\prime}{q}\,C_\mu\,\,\,\,\,.
\label{D}
\eea
is the scale invariant covariant derivative.
\par
In the actions (\ref{Ghilencea-2}) the index $a$ takes the values $1$ or $2$. 
The two constants $\sigma_1$ and $\sigma_2$ are adjusted such that the kinetic
terrm for the scalar field $\chi$ is of the canonical form  $+\,\frac{1}{2} \sqrt{-g}\, \dd_\mu \chi  \dd^\mu \chi$.
The desired values for  $\sigma_1$ and $\sigma_2$ are 
\bea
\sigma_1= {q^2}\left(\frac{1}{8} +\frac{3}{2}\xi\right) \,\,\,\,\,\,\,\,\,\,,\,\,\,\,\,\,\,\,\,\,
\sigma_2= \frac{1}{8}\,{q^2} \,\,\,\,.
\eea
With these two values, we have 
\be
S^{(1)}_{\text{scale}} = S^{(2)}_{\text{scale}} 
\ee
up to total derivative terms. 
\par
As explained in the next section, the two actions $S_1$ and  $S_2$ have a  natural interpretation 
in terms of Weyl geometry. However, the scale invariant actions in (\ref{Ghilencea-2}) lack
a geometrical interpretation. It is the aim of this note to provide such an interpretation.
\par
The paper is organised as follows: In the next section we quickly review Weyl geometry in connection 
with the two actions $S_1$ and  $S_2$. It is then followed by a subsection in which we recall that
the gauge kinetic term stems also from  Weyl geometry. In section 3, we present a method 
for scale invariance of gravity relying simultaneously on the gauge field $C_\mu$ and the scalar field 
$\varphi$. This has a natural interpretation in the context of  a generalised Weyl geometry as explained in section 4.
This constitutes the main result of this note.
\par
Sections 5 and 6 are dedicated to the scale invariant Standard Model as an application of the formalism.
In section 7, we obtain a curious result in the form of a duality transformations resembling scale transformations.
Finally, our conclusions and outlook are summarised in section 8.

\section{The usual Weyl geometry}

The actions $S_1$ and $S_2$ of the previous section  can be given a geometrical interpretation.
The Christoffel symbols
\be
\Gamma^\alpha_{\mu\nu}= \frac{1}{2}g^{\alpha\tau} \left(\dd_\mu g_{\tau\nu}+\dd_\nu g_{\tau\mu} - \dd_\tau g_{\mu\nu}\right)
\ee
transform under the scale symmetry $g_{\mu\nu}\longrightarrow e^{\sigma}\,g_{\mu\nu}$  as 
\be
\Gamma^\alpha_{\mu\nu} \longrightarrow \Gamma^\alpha_{\mu\nu} 
+\frac{1}{2}\left(\delta^\alpha_\mu\dd_\nu\sigma + \delta^\alpha_\nu\dd_\mu\sigma -g_{\mu\nu} g^{\alpha\tau} \dd_\tau\sigma\right)
\,\,\,\,\,.
\label{var-Chris}
\ee
One can then, with the help of the gauge field $C_\mu$, build the scale 
invariant symbols $\widetilde \Gamma^\alpha_{\mu\nu}$ as
\bea
\widetilde \Gamma^\alpha_{\mu\nu} &=&  \Gamma^\alpha_{\mu\nu}
- \frac{{q}'}{2}\left(\delta^\alpha_\mu C_\nu  + \delta^\alpha_\nu C_\mu
-g_{\mu\nu} g^{\alpha\sigma} C_\sigma \right)
\,\,,
\eea
As a consequence,  the scale invariant Riemann tensor  $\widetilde R^\mu_{\nu\rho\sigma}$ is defined as 
\be
\widetilde R^\mu_{\nu\rho\sigma}=\dd_\rho \widetilde\Gamma^\mu_{\nu\sigma} - \dd_\sigma \widetilde\Gamma^\mu_{\nu\rho} 
+ \widetilde\Gamma^\mu_{\rho\alpha}\widetilde \Gamma^{\alpha}_{\nu\sigma} -  \widetilde\Gamma^\mu_{\sigma\alpha} \widetilde\Gamma^{\alpha}_{\nu\rho}\,\,\,.
\label{Rie}
\ee
Finally, the scale invariant  Ricci tensor $\widetilde R_{\mu\nu}$ and  the Ricci scalar  $\widetilde R$ are
\be
\widetilde R_{\mu\nu}=\widetilde R^\alpha_{\mu\alpha\nu}\,\,\,\,\,\,,\,\,\,\,\,\,  \widetilde R=g^{\mu\nu}\widetilde R_{\mu\nu}
\,\,\,\,\, .
\label{Ricci}
\ee
\par
The expression of  the Ricci scalar   $\widetilde R$ is given by 
\be
 \widetilde R= R 
  + \left( 3{q'}\,\nabla_\mu C^\mu -\frac{3}{2}{{q}'}^2\,C_\mu C^\mu \right)\,\,\,\,\,.
\label{R-tilde}
\ee
Under the local transformation  $g_{\mu\nu}\longrightarrow e^{\sigma}\,g_{\mu\nu}$, the  Ricci scalar  $\widetilde R$ scales as
\be
\widetilde R \longrightarrow  e^{-\sigma}\, \widetilde R\,\,\,.
\ee
This suggests that the scale invariant theory should be given by the action
\bea
S_{\text{Weyl}}  &=& 
\int d^4x\sqrt{-g} \Bigg\{-\frac{\xi}{4\kappa} \widetilde {R}^2 \Bigg\}
\,\,\,.
\label{auxiliary-1}
\eea
This last action can be cast in the form
\bea
S_{\text{Weyl}}  &=& \xi\,\int d^4x\sqrt{-g}\,e^{-q\varphi}\Bigg\{- \widetilde {R} +\kappa\,e^{-q\varphi}\Bigg\}
\,\,\,.
\label{auxiliary}
\eea
The field $\varphi$  enters as a  non-propagating  auxiliary field whose
equations of motion is  
\be
e^{-q\varphi} = \frac{ \widetilde {R} }{2\kappa} \,\,\,.
\ee
The action $S_{\text{Weyl}}$ in  (\ref{auxiliary-1})
is obtained when using this equation of motion.
\par
Using the explicit expression of $\widetilde {R}$ as written (\ref{R-tilde}), the action (\ref{auxiliary}) is precisely
$S_2$ of the previous section. The elimination of the gauge field $C_\mu$ through its equation of motion gives then 
the action $S_1$.

\subsection{The gauge field kinetic term }

The scale invariant Ricci tensor $\widetilde R_{\mu\nu}$  as defined in (\ref{Ricci}) 
is not symmetric in its indices. Its anti-symmetric part is given by
\bea
\widetilde R_{[\mu\nu]} =\frac{1}{2}\left(\widetilde R_{\mu\nu} - \widetilde R_{\nu\mu}\right)
=- {q'}\,C_{\mu\nu} \,\,\,\,\,\,\,\,\,,\,\,\,\,\,\,\,\,\, C_{\mu\nu} = \dd_\mu C_\nu -\dd_\nu C_\mu
\,\,\,\,.
\eea
Therefore, one could add to (\ref{auxiliary-1}) the scale invariant action
\bea
S_{\text{add}} &=& \frac{1}{{q'}^2} \int d^4x\sqrt{-g}\Bigg\{-\frac{1}{4}\widetilde R_{[\mu\nu]}\,\widetilde R^{[\mu\nu]}\Bigg\}
\nonumber \\
&=& \int d^4x\sqrt{-g}\Bigg\{-\frac{1}{4}C_{\mu\nu}C^{\mu\nu} \Bigg\} 
\,\,\,.
\eea
This known observation assigns a geometric origin to the kinetic term of the gauge field $C_\mu$.

\section{An alternative way}

It is now clear that scale symmetry of gravity requires a scalar field $\varphi$ and a gauge field 
$C_\mu$.
Under the scale transformations
\bea
g_{\mu\nu} &\longrightarrow& e^{\sigma(x)}\, g_{\mu\nu} \,\,\,\,\,,
\nonumber \\
\varphi &\longrightarrow &  \varphi + \frac{1}{q}\, \sigma\,\,\,\,\,,
\nonumber \\
C_\mu &\longrightarrow&  C_\mu + \frac{1}{{q}'}\,\dd_\mu \sigma\,\,\,\,\,
\label{shift-0}
\eea
one obtains  the changes
\bea 
R  &\longrightarrow &  e^{-\sigma(x)}\,\left(R -\Delta R\right) \,\,\,\,\,,
\nonumber \\
  3q\,\nabla^2\varphi -\frac{3}{2}q^2\,\dd_\mu\varphi \dd^\mu\varphi
&\longrightarrow &
 e^{-\sigma(x)}\,\left( 3q\,\nabla^2\varphi -\frac{3}{2}q^2\,\dd_\mu\varphi \dd^\mu\varphi 
+\Delta R\right) \,\,,
\nonumber \\
 3{q'}\,\nabla_\mu C^\mu -\frac{3}{2}{{q}'}^2\,C_\mu C^\mu
&\longrightarrow &
 e^{-\sigma(x)}\,\left(  3{q'}\,\nabla_\mu C^\mu -\frac{3}{2}{{q}'}^2\,C_\mu C^\mu
+ \Delta R \right) \,\,,
\eea
where 
\be
\Delta R =  3\nabla^2\sigma + \frac{3}{2}\dd_\mu\sigma \dd^\mu\sigma \,\,\,\,\,.
\ee
Notice that the cancelation of the variation $\Delta R$ coming from the Ricci scalar $R$ can
be achieved by including either the scalar field $\varphi$ or the gauge field $C_\mu$ or both.
It is this last option that we adopt. 
\par
The scale invariant gravitational theory we propose is given by the action
\bea
S_\varepsilon  &=& 
\xi\,\int d^4x\sqrt{-g}\,e^{-q\varphi}\Bigg\{- R - \varepsilon\left( 3q\,\nabla^2\varphi -\frac{3}{2}q^2\,\dd_\mu\varphi \dd^\mu\varphi\right)
\nonumber \\
 &-& (1-\varepsilon)\left( 3{q'}\,\nabla_\mu C^\mu -\frac{3}{2}{{q}'}^2\,C_\mu C^\mu \right)+\kappa\,e^{-q\varphi}\Bigg\}
\,\,\,\,\,.
\label{epsilon-action}
\eea
Here $\varepsilon$ is a constant. In particular, the two actions $S_1$ and $S_2$ are obtained, respectively,  for the special values
$\varepsilon=1$ and $\varepsilon=0$.
\par
This last  action can be written as
\bea
S_\varepsilon  &=& 
\xi\,\int d^4x\sqrt{-g}\,e^{-q\varphi}\Bigg\{- R 
  - \left( 3{q'}\,\nabla_\mu C^\mu -\frac{3}{2}{{q}'}^2\,C_\mu C^\mu \right)+\kappa\,e^{-q\varphi} \Bigg\}
\nonumber \\
&-& \frac{3}{2}q^2\varepsilon\xi \int d^4x\sqrt{-g}\,e^{-q\varphi} \,\cal D_\mu\varphi \cal D^\mu\varphi  
\,\,\,\,\,.
\label{epsilon-action-1}
\eea
The total derivative term we  discarded is $-3\xi \varepsilon q\int d^4x\sqrt{-g}\,\nabla_\mu\left(e^{-q\varphi}\cal D^\mu\varphi\right)$ and
the covariant derivative ${\cal D}_\mu \varphi$ is given in (\ref{D}).
Furthermore, we see that the action (\ref{epsilon-action-1}) is 
\be
S_\varepsilon  =  S_2  - \frac{3}{2}q^2\varepsilon\xi \int d^4x\sqrt{-g}\,e^{-q\varphi} \,\cal D_\mu\varphi \cal D^\mu\varphi  
\,\,\,\,\,.
\ee
This is the sum of two scale invariant parts.
\par
In a similar way, one can show that the action (\ref{epsilon-action}) can also be written as
\bea
S_\varepsilon  &=& 
\xi\,\int d^4x\sqrt{-g}\,e^{-q\varphi}\Bigg\{- R - \left( 3q\,\nabla^2\varphi -\frac{3}{2}q^2\,\dd_\mu\varphi \dd^\mu\varphi\right)
\nonumber \\
&+& \frac{3}{2}q^2 \left(1-\varepsilon\right)\xi \int d^4x\sqrt{-g}\,e^{-q\varphi} \,\cal D_\mu\varphi \cal D^\mu\varphi  
\,\,\,\,\,.
\label{epsilon-action-2}
\eea
This up to the total derivative term ${3}q \left(1-\varepsilon\right)\xi \int d^4x\sqrt{-g}\,\nabla_\mu\left(e^{-q\varphi}\cal D^\mu\varphi\right)$.
Hence, we have also
\be
S_\varepsilon  =  S_1  + \frac{3}{2}q^2 \left(1-\varepsilon\right)\xi \int d^4x\sqrt{-g}\,e^{-q\varphi} \,\cal D_\mu\varphi \cal D^\mu\varphi  
\,\,\,\,\,.
\ee
Therefore, the action proposed in (\ref{epsilon-action}) is the parent theory for $S_1$ and   $S_2$. These 
are now equivalent up the a scale invariant term proportional to $\int d^4x\sqrt{-g}\,e^{-q\varphi} \,\cal D_\mu\varphi \cal D^\mu\varphi$. 

\par
We notice that a positive canonical kinetic term 
\be
+\,\frac{1}{2} \sqrt{-g}\, \dd_\mu \chi  \dd^\mu \chi \,\,\,,
\ee
where the scalar field $\chi$ is as defined in 
(\ref{chi}), is obtained for 
\be
\varepsilon=-\frac{1}{12\xi}\,\,\,\,\,.
\label{epsilon}
\ee

\section{A general Weyl geometry}

The variation, due to the rescaling of the metric, of the Christoffel symbols as given in (\ref{var-Chris})
can be cancelled by taking the scale invariant symbols $\widetilde \Gamma^\alpha_{\mu\nu}$ to be given by
\bea
\widetilde \Gamma^\alpha_{\mu\nu} &=& \Gamma^\alpha_{\mu\nu} 
-\frac{{q}}{2}\omega\left(\delta^\alpha_\mu \dd_\nu\varphi  + \delta^\alpha_\nu \dd_\mu\varphi
-g_{\mu\nu} g^{\alpha\sigma} \dd_\sigma\varphi \right)
\nonumber\\
&-& \frac{{q}'}{2}\left(1-\omega\right)\left(\delta^\alpha_\mu C_\nu  + \delta^\alpha_\nu C_\mu
-g_{\mu\nu} g^{\alpha\sigma} C_\sigma \right)
\,\,,
\label{Weyl-Gamma-0}
\eea
where $\omega$ is a constant. This can be also written as 
\bea
\widetilde \Gamma^\alpha_{\mu\nu} &=& \Gamma^\alpha_{\mu\nu} 
 - \frac{{q}'}{2}\left(\delta^\alpha_\mu C_\nu  + \delta^\alpha_\nu C_\mu
-g_{\mu\nu} g^{\alpha\sigma} C_\sigma \right)
\nonumber\\
 &-& \frac{{q}}{2}\omega\left(\delta^\alpha_\mu {\cal D}_\nu\varphi  + \delta^\alpha_\nu {\cal D}_\mu\varphi
-g_{\mu\nu} g^{\alpha\sigma} {\cal D}_\sigma\varphi \right)
\,\,.
\label{Weyl-Gamma}
\eea
The scale invariant Riemman tensor $\widetilde R^\mu_{\nu\rho\sigma}$ is still as defined in
(\ref{Rie}) and  the scale invariant  Ricci tensor $\widetilde R_{\mu\nu}$ and  the Ricci scalar  $\widetilde R$ are
as given in (\ref{Ricci}).
\par
An explicit calculation gives 
\bea
\widetilde R &=& R +3q\omega \nabla_\mu {\cal D}^\mu\varphi + 3q^\prime \nabla_\mu C^\mu
\nonumber\\
&-&\frac{3}{2}\left(q^2\omega^2 {\cal D}_\mu\varphi {\cal D}^\mu\varphi + 2\omega qq^\prime {\cal D}_\mu\varphi C^\mu 
+{q^\prime }^2 C_\mu C^\mu\right)
\,\,\,.
\label{Weyl-R}
\eea
This allows us to write the two  useful relations
\bea
e^{-q\varphi}\,\widetilde R &=& e^{-q\varphi}\left[ R + 3q^\prime \nabla_\mu C^\mu 
-\frac{3}{2}{q^\prime }^2 C_\mu C^\mu + \frac{3}{2}q^2\omega\left(2-{\omega}{}\right)
\cal D_\mu\varphi \cal D^\mu\varphi\right]
\nonumber\\
&+& 3q\omega \nabla_\mu \left(e^{-q\varphi}\, \cal D^\mu\varphi\right) 
\,\,\,
\label{rel-1}
\eea
and 
\bea
e^{-q\varphi}\,\widetilde R &=& e^{-q\varphi}\left[ R + 3q \nabla^2 \varphi
-\frac{3}{2}{q}^2\dd_\mu\varphi \dd^\mu\varphi  - \frac{3}{2}q^2\left(1-{\omega}{}\right)^2
\cal D_\mu\varphi \cal D^\mu\varphi\right]
\nonumber\\
&-& 3q\left(1-\omega\right) \nabla_\mu \left(e^{-q\varphi}\, \cal D^\mu\varphi\right) 
\,\,\,.
\label{rel-2}
\eea
\par
By making the identification
\be
\varepsilon =\omega\left(2-\omega\right) \,\,\,
\label{id}
\ee
and using the two relations (\ref{rel-1}) and  (\ref{rel-2}), 
the action  $S_\varepsilon$ in  (\ref{epsilon-action-1}) or in  (\ref{epsilon-action-2})  can be both written as 
\bea
S_{\varepsilon}  &=& 
\xi\,\int d^4x\sqrt{-g}\,e^{-q\varphi}\Bigg\{- \widetilde {R} +\kappa\,e^{-q\varphi}\Bigg\}
 \,\,\,.
\label{S-epsilon}
\eea
This is up to total derivative terms. In this action, the term   $e^{-q\varphi}\,\widetilde R$
is as given in (\ref{rel-1}) or  (\ref{rel-2}).
\par
In the action (\ref{S-epsilon}), the scalar field $\varphi$ is no longer a Lagrange multiplier
as  $\widetilde {R}$ contains derivatives of this field.
The kinetic term  $+\,\frac{1}{2} \sqrt{-g}\, \dd_\mu \chi  \dd^\mu \chi$,  for the scalar field $\chi$ of (\ref{chi}), stems 
from the Ricci scalar $\widetilde {R}$ when $\varepsilon=-\frac{1}{12\xi}$. 
\par
The anti-symmetric part of the scale invariant Ricci tensor  $\widetilde R_{\mu\nu}$  as defined in (\ref{Ricci}),
but with the scale invariant symbols $\widetilde \Gamma^\alpha_{\mu\nu}$ now given in (\ref{Weyl-Gamma}), is found 
to  be given as 
\bea
\widetilde R_{[\mu\nu]} =\frac{1}{2}\left(\widetilde R_{\mu\nu} - \widetilde R_{\nu\mu}\right)
=- {q'}\left(1-\omega\right)\,C_{\mu\nu} 
\,\,\,\,.
\label{anti-Ric}
\eea
We could supplement the gauge field $C_\mu$ with a kinetic term by including the additional action 
\bea
S_{\text{add}} &=& \frac{1}{{q'}^2\left(1-\omega\right)^2} \int d^4x\sqrt{-g}\Bigg\{-\frac{1}{4}\widetilde R_{[\mu\nu]}\,\widetilde R^{[\mu\nu]}\Bigg\}
\nonumber \\
&=& \int d^4x\sqrt{-g}\Bigg\{-\frac{1}{4}C_{\mu\nu}C^{\mu\nu} \Bigg\} 
\,\,\,.
\eea
We are of course assuming that $\omega \ne 1$ (or, according to (\ref{id}), $\varepsilon \ne 1$). If $\omega=1$, then 
the gauge field  $C_\mu$ does not take part in the construction  of $\widetilde \Gamma^\alpha_{\mu\nu}$ as can be seen from (\ref{Weyl-Gamma-0}).
\par
In summary, a scale invariant theory of gravity could be described by the geometric action
\bea
S_{{\text {scale}}} &=&  S_{\varepsilon} +  S_{\text{add}}
\nonumber \\
&=& \xi\,\int d^4x\sqrt{-g}\,e^{-q\varphi}\Bigg\{- \widetilde {R} +\kappa\,e^{-q\varphi}\Bigg\}
\nonumber \\
&+&  
\frac{1}{{q'}^2\left(1-\omega\right)^2} \int d^4x\sqrt{-g}\Bigg\{-\frac{1}{4}\widetilde R_{[\mu\nu]}\,\widetilde R^{[\mu\nu]}\Bigg\}
\,\,\,.
\label{action-scale}
\eea
The expressions of $\widetilde {R}$ and $\widetilde R_{[\mu\nu]}$ are given, respectively, in (\ref{Weyl-R})
and (\ref{anti-Ric}). Notice that $\left(1-\omega\right)^2 =  \left(1-\varepsilon \right)$, as can be seen from (\ref{id}),
and for a positive kinetic term for the scalar field $\chi$, we take $\varepsilon=-\frac{1}{12\xi}$. 
\par
The action (\ref{action-scale}), when unpacked, is also given in ref.\cite{sm2} where different terms were
introduced by 'hand'. Here we provide a geometric origin to the scale invariant theory.

\section{Application: scale invariance of the Standard Model coupled to gravity }

In this section we supplement the Standard Model with the scalar field $\varphi$ and the gauge field 
$C_\mu$ and couple it to gravity in a scale  invariant manner.
The scale transformations of the fields of the Standard Model are 
\bea
H &\,\longrightarrow \,&  e^{-\frac{1}{2}\sigma(x)}\, H \,\,\,\,\,\,,
\nonumber \\
\psi^{(a)} &\,\longrightarrow \,&   e^{-\frac{3}{4}\sigma(x)}\, \psi^{(a)} \,\,\,\,\,\,,
\nonumber \\
A_\mu^{(i)}  &\,\longrightarrow \,& A_\mu^{(i)} \,\,\,\,\,\,, 
\label{rescale}
\eea
where $H$, $\psi^{(a)}$ and $A_\mu^{(i)}$ represent, respectively, the Higgs doublet, any fermion field and any gauge field
in the Standard Model. 
\par
The scale invariant Standard Model coupled to gravity could be described by the action
\bea
S &=& 
\xi\,\int d^4x\sqrt{-g}\,\left(e^{-q\varphi} + \zeta\,H^\dagger H\right)  \Bigg\{- R - \varepsilon\left( 3q\,\nabla^2\varphi -\frac{3}{2}q^2\,\dd_\mu\varphi \dd^\mu\varphi\right)
\nonumber \\
 &-& (1-\varepsilon)\left( 3{q'}\,\nabla_\mu C^\mu -\frac{3}{2}{{q}'}^2\,C_\mu C^\mu \right)+\kappa\,e^{-q\varphi} \Bigg\}
\nonumber \\
&+& \int d^4x\sqrt{-g}\Bigg\{-\frac{1}{4}C_{\mu\nu}C^{\mu\nu} -\frac{\gamma}{2}\,C_{\mu\nu}B^{\mu\nu} 
\Bigg\}
\nonumber \\
&+& \int d^4x\sqrt{-g}\,\Bigg\{  \left(D_\mu H\right)^\dagger\left(D^\mu H\right)
 - {\lambda}\left(H^\dagger H\right)^2 + \dots \,\,\,\,\,\, \Bigg\} \,\,\,\,\,\,.
\label{SM-scale-action}
\eea
We have included a mixing term, with a parameter $\gamma$,  between the field strength $C_{\mu\nu}$ and the $U(1)$ field
strength $B_{\mu\nu} = \dd_\mu B_\nu -  \dd_\nu B_\mu$ of the Standard Model.  The last contribution in this action
is the Standard Model action with the Higgs mass set to zero\footnote{In the absence of gravity and with a mass term for the Higgs field, 
this theory represents a St\"uckelberg extension of the Standard Model (see, for instance, \cite{ex1,ex2}).} 
and the dots stand for the rest of the Standard Model Lagrangian (written in a covariant form).
The non-minimal coupling of the Higgs field to the scalar curvature, with parameter $\zeta$,  is permitted by the scale symmetry.
Of course, we take  $\varepsilon=-\frac{1}{12\xi}$, as in (\ref{epsilon}),  for a positive kinetic term for the scalar field $\chi$ in (\ref{chi}).
\par
The gauge covariant derivative acting on the 
complex doublet $H$ is now given by\footnote{We could have taken
$D_\mu H = \left[\dd_\mu - \frac{i}{2}{g}'B_\mu - i {g}W_\mu  + \frac{\tau}{2}{q}'C_\mu +\frac{1}{2} (1-\tau)q\dd_\mu\varphi \right]H$.
However, if the real parameter $\tau$ is different from $1$ then the theory is not renormalisable in the absence of gravity.} 
\bea
D_\mu H = \left[\dd_\mu - \frac{i}{2}{g}'B_\mu - i {g}W_\mu  + \frac{1}{2}{q}'C_\mu \right]H
\,\,\,\,\,.
\eea
The $SU(2)$ and the $U(1)$ gauge couplings are, respectively, $g$ and ${g}'$.
\par
Since, the gauge fields do not scale, all the gauge kinetic terms in the Standard Model Lagrangian are scale 
invariant. The kinetic terms corresponding to the fermions as well as the Yukawa interaction terms are also scale invariant (see appendix A). 
We assume that the fermions are not charged with respect to the additional gauge field $C_\mu$.

\section{A gauge choice : The Higgs non-minimaly coupled to gravity}

The extra scalar field transforms as $\varphi \longrightarrow \varphi +\frac{1}{q}\,{\sigma(x)}$. This calls for 
the obvious gauge fixing 
\be
\varphi =0 \,\,\,\,\,.
\ee
In this gauge, the action (\ref{SM-scale-action}) becomes
\bea
S &=& 
\int d^4x\sqrt{-g}\, \Bigg\{-\xi\, \left(1 + \zeta\,H^\dagger H\right) R + \xi \kappa
\nonumber \\
 &-&\xi\,\zeta\, (1-\varepsilon)\,H^\dagger H \left[ 3{q'}\,\nabla_\mu C^\mu -\frac{3}{2}{{q}'}^2\,C_\mu C^\mu \right] \Bigg\}
\nonumber \\
&+& \int d^4x\sqrt{-g}\Bigg\{-\frac{1}{4}C_{\mu\nu}C^{\mu\nu} -\frac{\gamma}{2}\,C_{\mu\nu}B^{\mu\nu} 
+\frac{1}{2}m^2\,C_\mu C^\mu \, \Bigg\}
\nonumber \\
&+& \int d^4x\sqrt{-g}\,\Bigg\{  \left(D_\mu H\right)^\dagger\left(D^\mu H\right) - M^2\,H^\dagger H
 - {\lambda}\left(H^\dagger H\right)^2 + \dots \,\,\,\,\,\, \Bigg\} \,\,\,\,\,\,.
\label{gauge-fixed-action}
\eea
We have discarded the total derivative  -$3\xi{{q}'}\left(1-\varepsilon \right)\int d^4x\sqrt{-g}\,\nabla_\mu C^\mu$  and made the identification
\bea
M^2 & = &-\zeta \xi\kappa 
\nonumber \\
m^2 & = & 3\xi{{q}'}^2\left(1-\varepsilon \right)\,\,\,\,\,\,.
\eea
We notice that the Higgs field $H$ acquires a mass upon the breaking of the scale symmetry.
Similarly, the gauge field $C_\mu$ becomes massive. The two constants $\xi$ and $\zeta$ should satisfy the relation
\be
\xi\, \left(1 + \zeta\,v^2\right)=\frac{{M^2_{\text P}}}{2} \,\,\,\,,
\label{vev}
\ee
where  ${M_{\text P}}$ is the reduced Planck mass and $v^2=-\frac{{M^2}}{2\lambda}$ is the Higgs expectation value (assuming that 
$M^2$ is negative). This relation means that the Planck mass is generated by the Higgs expectation value.
\par
The action (\ref{gauge-fixed-action}) when the gauge field $C_\mu$ is absent and\footnote{This relation excludes a cosmological constant term
and gives the  Higgs field the potential $ {\lambda}\left(H^\dagger H  -v^2\right)^2$ with $v^2=-\frac{{M^2}}{2\lambda}$.}
\be
\xi \kappa + \lambda\left(\frac{{M^2}}{2\lambda}\right)^2 =0 \,\,\,\,
\label{cond}
\ee
is exactly that studied in the context of inflation based on a non-minimally coupled Higgs field to gravity \cite{kaiser, bezrukov1,bezrukov2}.  
\par
Indeed, the coupling between the Higgs field and the Ricci scalar $R$ can be absorbed by 
expressing the action (\ref{gauge-fixed-action}) in the Einstein frame. This is achieved by rescaling
the metric as
\bea
g_{\mu\nu} \longrightarrow e^{-\alpha\,\rho}  \,g_{\mu\nu}
\,\,\,\,,
\eea
where $\alpha$ is a constant (to be fixed later) and the scalar field $\rho(x)$
is such that 
\be
 e^{-\alpha\,\rho} \,{\left(1 + \zeta\,H^\dagger H\right) = \frac{M^2_{\text P}}{2\xi}}\,\,\,\,.
\ee 
The field $\rho(x)$ will be identified with the inflaton field \cite{me}.
\par
In the unitary gauge in which the Higgs doublet $H$ is given by
\be
H=\left(
\begin{array}{c}
0\\
h(x)
\end{array}\right)
\ee
the action (\ref{gauge-fixed-action}), in the Einstein frame,  takes the form
\bea
S &=& 
\int d^4x\sqrt{-g}\, \Bigg\{-\frac{M^2_{\text P}}{2}\, R + 
\frac{3}{4}\alpha^2 {M^2_{\text P}}
\Bigg [1 + \frac{1}{6}   \frac{1}{\zeta\xi}\frac{1}{\left(1-\frac{2\xi}{M^2_{\text P}}\, e^{-\alpha\rho}\right)}\Bigg]
g^{\mu\nu}\dd_\mu\rho \dd_\nu\rho 
\nonumber \\
&-& V\left(\rho\right)+\dots \Bigg \}
\,\,\,\,.
\eea
We have maintained only the terms relevant to inflation and the potential $V\left(\rho\right)$
is given by
\bea
V\left(\rho\right) =  e^{-2\alpha\rho}\Bigg\{
\frac{\lambda}{\zeta^2} \left(\frac{M^2_{\text P}} {2\xi}\right)^2 \Bigg[e^{\alpha\rho}-1\Bigg]^2
-\left[\xi \kappa + \lambda\left(\frac{{M^2}}{2\lambda}\right)^2\right] \Bigg \} \,\,\,\,.
\eea
The relation in (\ref{vev}) has been used in obtaining this potential.
\par
When the condition (\ref{cond}) holds and 
\bea
\alpha = \sqrt{\frac{2}{3}}\frac{1}{M_{\text P}} \,\,\,\,\,\,\,\,\,\,,\,\,\,\,\,\,\,\,\,\,
\frac{2\xi}{M^2_{\text P}}\, e^{-\alpha\rho} \ll 1\,\,\,\,\,\,\,\,\,\,,\,\,\,\,\,\,\,\,\,\,
\frac{1}{6}   \frac{1}{\zeta\xi} \ll 1
\eea
inflation is driven by the scalar field $\rho$ \cite{me} with approximate kinetic term $\frac{1}{2}g^{\mu\nu}\dd_\mu\rho \dd_\nu\rho$.
In this limit the potential is flat (slow-roll condition) and is dominated by a cosmological constant equal to  
$\frac{\lambda}{\zeta^2} \left(\frac{M^2_{\text P}} {2\xi}\right)^2$.

\section{ A duality transformation}

We will consider the same action as in (\ref{SM-scale-action}) supplemented with a cosmological constant $\Lambda$
and  mass terms for the complex doublet $H$ and the extra field $\varphi$. More precisely, we take the action
\bea
S &=& 
\xi\,\int d^4x\sqrt{-g}\,\left(e^{-q\varphi} + \zeta\,H^\dagger H\right)  \Bigg\{- R - \varepsilon\left( 3q\,\nabla^2\varphi -\frac{3}{2}q^2\,\dd_\mu\varphi \dd^\mu\varphi\right)
\nonumber \\
 &-& (1-\varepsilon)\left( 3{q'}\,\nabla_\mu C^\mu -\frac{3}{2}{{q}'}^2\,C_\mu C^\mu \right)+\kappa\,e^{-q\varphi} \Bigg\}
\nonumber \\
&+& \int d^4x\sqrt{-g}\Bigg\{-\frac{1}{4}C_{\mu\nu}C^{\mu\nu} -\frac{\gamma}{2}\,C_{\mu\nu}B^{\mu\nu}  \Bigg\}
\nonumber \\
&+& \int d^4x\sqrt{-g}\,\Bigg\{  \left(D_\mu H\right)^\dagger\left(D^\mu H\right)
 - {\lambda}\left(H^\dagger H\right)^2 + \dots \,\,\,\,\,\, \Bigg\} \,\,\,\,\,\,
\nonumber \\
&+& \int d^4x\sqrt{-g}\,\Bigg\{-\Lambda -M^2\,H^\dagger H - \frac{1}{2}\,m_\chi^2\,e^{-q\varphi} \Bigg\} \,\,\,\,\,\,.
\label{duality-action}
\eea
Of course, the additional last term in this theory breaks scale invariance and we would like to
see to what extent this  scale symmetry is broken. 
\par
Indeed, the last  term  transforms 
under the scale symmetry $g_{\mu\nu} \longrightarrow e^{\sigma}\, g_{\mu\nu}$, $ H \longrightarrow   e^{-\frac{1}{2}\sigma}\, H$ and 
$ \varphi \longrightarrow  \varphi + \frac{1}{q}\,\sigma$ as
\bea
 \sqrt{-g}\,\Bigg\{-\Lambda - M^2\,H^\dagger H -\frac{1}{2}\, m_\chi^2\,e^{-q\varphi} \Bigg\}\, \longrightarrow\, 
\sqrt{-g}\,\Bigg\{-\Lambda\,e^{2\sigma} -M^2\,H^\dagger H\,e^{\sigma} -\frac{1}{2}\, m_\chi^2\,e^{-q\varphi} \,e^{\sigma}  \Bigg\}
\,\,.
\nonumber \\
\eea
Let us now demand that 
\be
\Bigg\{-\Lambda\,e^{2\sigma} - M^2\,H^\dagger H\,e^{\sigma}  -\frac{1}{2}\, m_\chi^2\,e^{-q\varphi} \,e^{\sigma}  \Bigg\}= 
\Bigg\{-\Lambda - M^2\,H^\dagger H  -\frac{1}{2}\, m_\chi^2\,e^{-q\varphi}   \Bigg\}\,\,\,\,\,\,.
\ee
This last relation is satisfied if $\sigma$ is given by
\be
e^{\sigma} = -1 -\frac{1}{\Lambda}\left({M^2}\, H^\dagger H +\frac{1}{2}\, m_\chi^2\,e^{-q\varphi}\right) \,\,\,\,\,\,.
\label{sigma}
\ee
To summarise, the action (\ref{duality-action}) remains unchanged under the transformations
\bea
g_{\mu\nu} & \longrightarrow &  e^{\sigma}\, g_{\mu\nu}\,\,\,\,\,,
\nonumber \\
\varphi &\longrightarrow &  \varphi + \frac{1}{q}\, \sigma\,\,\,\,\,,
\nonumber \\
C_\mu &\longrightarrow&  C_\mu + \frac{1}{{q}'}\,\dd_\mu \sigma\,\,\,\,\,,
\nonumber \\
 H &\longrightarrow &   e^{-\frac{1}{2}\sigma}\, H\,\,\,\,\,,
\nonumber \\
\psi^{(a)} &\,\longrightarrow \,&   e^{-\frac{3}{4}\sigma}\, \psi^{(a)} \,\,\,\,\,\,,
\nonumber \\
A_\mu^{(i)}  &\,\longrightarrow \,& A_\mu^{(i)} \,\,\,\,\,\,. 
\eea
The scale factor $\sigma$ is a function of the two fields $H$ and $\varphi$  as given in (\ref{sigma}). This is a duality transformation since  $\sigma$ is not an arbitrary 
function but depends on the  fields  $H$ and $\varphi$ of the theory. As usual, $\psi^{(a)} $ and $A_\mu^{(i)}$
are the not-shown fermions and gauge fields of the Standard Model Lagrangian. This duality transformation assumes that the cosmological
constant $\Lambda$ is different from zero and is valid in a domain  where 
\be
 -1 -\frac{1}{\Lambda}\left({M^2}\, H^\dagger H +\frac{1}{2}\, m_\chi^2\,e^{-q\varphi}\right)  \ge 0\,\,\,\,\,\,. 
\ee

\section{Conclusions}

A scale invariant theory of gravity necessitates, beside the metric tensor field $g_{\mu\nu}$, a scalar field $\varphi$ and a gauge field $C_\mu$.
We have in this note assembled these three fields together and constructed a scale invariant Christoffel symbols $\widetilde \Gamma^\alpha_{\mu\nu}$
as given in (\ref{Weyl-Gamma}).  The scalar field $\varphi$ is then a physical field having a positive kinetic term.
As an application of the formalism, we have coupled the Standard Model to gravity in a scale invariant manner. The widely investigated 
subject of the non-minimally coupled  Standard Model to gravity is obtained as a consequence of the breaking of this scale symmetry.
However, the theory contains an extra gauge field $C_\mu$. It would be of great interest to explore the issues brought by the presence
of this gauge field in the context of inflation in the early Universe.
\par
The analyses carried out in this article could be easily generalised to the case of $N$ scalar fields $\varphi^i$ with $i=1,\dots,N$.
The scale  invariant Christoffel symbols $\widetilde \Gamma^\alpha_{\mu\nu}$ are then written as 
\bea
\widetilde \Gamma^\alpha_{\mu\nu} &=& \Gamma^\alpha_{\mu\nu} 
 - \frac{{q}'}{2}\left(\delta^\alpha_\mu C_\nu  + \delta^\alpha_\nu C_\mu
-g_{\mu\nu} g^{\alpha\sigma} C_\sigma \right)
\nonumber\\
 &-& \frac{{v_i}}{2}\left(\delta^\alpha_\mu {\cal D_\nu}\varphi^i  + \delta^\alpha_\nu {\cal D}_\mu\varphi^i
-g_{\mu\nu} g^{\alpha\sigma} {\cal D}_\sigma\varphi^i \right)
\,\,,
\label{Weyl-Gamma-1}
\eea
where  ${\cal D}_\mu \varphi^i = \dd_\mu \varphi^i -\frac{q^\prime}{q_i}\,C_\mu$ and 
$v_i$ and $q_i$ are constants. Each scalar field transforms as $\varphi^i\longrightarrow \varphi^i+\frac{1}{q_i}\sigma$.
\par
The Ricci scalar constructed out of the  scale invariant Christoffel symbols $\widetilde \Gamma^\alpha_{\mu\nu}$ in (\ref{Weyl-Gamma-1})
is then given by
\bea
\widetilde R &=& R +3v_i \nabla_\mu {\cal D}^\mu\varphi^i + 3q^\prime \nabla_\mu C^\mu
\nonumber\\
&-&\frac{3}{2}\left(v_iv_j{\cal D}_\mu\varphi^i {\cal D}^\mu\varphi^j + 2v_i q^\prime {\cal D}_\mu\varphi^i C^\mu 
+{q^\prime }^2 C_\mu C^\mu\right)
\,\,\,.
\label{Weyl-R-1}
\eea
Under the local transformation  $g_{\mu\nu}\longrightarrow e^{\sigma}\,g_{\mu\nu}$, the  Ricci scalar  $\widetilde R$ scales as
$\widetilde R \longrightarrow  e^{-\sigma}\, \widetilde R$. 
\par
Similarly, the anti-symmetric part of the scale invariant Ricci tensor $\widetilde R_{\mu\nu}$
is 
\bea
\widetilde R_{[\mu\nu]} =\frac{1}{2}\left(\widetilde R_{\mu\nu} - \widetilde R_{\nu\mu}\right)
=- {q'}\left(1-\frac{v_i}{q_i}\omega\right)\,C_{\mu\nu} 
\,\,\,\,.
\label{anti-Ric-1}
\eea
A sum over the repeated index $i$ is understood.
The scale invariant theory is described  by the action
\bea
S_{\text{scale}}  
 &=& \xi\,\int d^4x\sqrt{-g}\,e^{-\frac{q_i\varphi^i}{N}}\Bigg\{- \widetilde {R} +\kappa\,e^{-\frac{q_j\varphi^j}{N}}\Bigg\}
\nonumber \\
&+& \frac{1}{{q'}^2\left(1-\frac{v_i}{q_i}\right)^2} \int d^4x\sqrt{-g}\Bigg\{-\frac{1}{4}\widetilde R_{[\mu\nu]}\,\widetilde R^{[\mu\nu]}\Bigg\}
\,\,\,.
\eea
The expressions of $\widetilde R$ and $\widetilde R_{[\mu\nu]}$ are as given in (\ref{Weyl-R-1}) and (\ref{anti-Ric-1}), respectively.
\par
Finally, we have shown in section 5, that  there is a kind of a duality transformation in the  theory written in (\ref{duality-action}).
This duality holds  only  in the presence of a cosmological constant. An extension of the present study would be a further exploration of the
consequences of this duality transformation.

\newpage

\appendix

\section{Scale invariance of a fermionic kinetic term and a Yukawa coupling}

The kinetic part of a Lagrangian for  a fermion field $\psi$ is given by 
\bea
L_{\mathrm{fermion}} &=& i \sqrt{-g}\,\bar\psi \gamma^a E_a^\mu D_\mu\psi \,\,\,\,,
\eea
where
\bea
D_\mu\psi &=& \left(\dd_\mu + \frac{1}{2} \omega^{bc}\Sigma_{bc}+\dots\right) \psi \,\,\,\,.
\eea
The dots in $ D_\mu\psi$ stand for possible coupling to gauge fields. Here we use the notation
\bea
g_{\mu\nu}= \eta_{ab}\,e^a_\mu e^b_\mu \,\,\,\,\,\,\,\,,\,\,\,\,\,\,\,\, 
e^a_\mu E^\mu_b =\delta^a_b \,\,\,\,\,\,\,\,,\,\,\,\,\,\,\,\, 
e^a_\mu E^\nu_a =\delta^\nu_\mu \,\,\,\,\,\,.
\eea 
The vielbeins scale as $e^a_\mu\longrightarrow e^{\sigma/2}\, e^a_\mu$ and their inverses as  $E_a^\mu\longrightarrow e^{-\sigma/2}\, E_a^\mu$.
Latin indices are raised and lowered with the flat metric $\eta_{ab}$ and its inverse $\eta^{ab}$. The spin connection is given by 
\be
\omega^a_{\mu b} = - E^\nu_b\left(\dd_\mu e^a_\nu - \Gamma^\alpha_{\mu\nu}\,e^a_\alpha\right)\,\,\,\,\,\,.
\ee
We ave also 
\be
\Sigma_{ab} =\frac{1}{4}\left[\gamma_a\,,\,\gamma_b \right] 
\,\,\,\,\,\,\,\,,\,\,\,\,\,\,\,\, 
\gamma^a\gamma^b + \gamma^b\gamma^a = 2\eta^{ab}\,\,\,\,\,.
\label{Dirac}
\ee
The spin connection transforms under the scaling symmetry $g_{\mu\nu} \longrightarrow e^{\sigma}\, g_{\mu\nu}$  as
\be
\omega^a_{\mu b}  \longrightarrow \omega^a_{\mu b} 
+ \frac{1}{2}\left( E^\nu_b\,e^a_\nu - \eta^{ad} \,E^\nu_d \, e^c_\mu \,\eta_{cb} \right)\dd_\nu\sigma
\,\,\,\,\,\,.
\ee
The covariant derivative then changes as
\bea
D_\mu \psi  \longrightarrow e^{-3\sigma/4}\,\left[ D_\mu \psi  -\frac{3}{4}\,\dd_\mu\sigma \,\psi 
+  \frac{1}{4}\, \left(\eta^{bc} E^\nu_c\,e^a_\nu - \eta^{ad} \,E^\nu_d \, e^b_\mu \right)\dd_\nu\sigma\,\Sigma_{ab} \,\psi \right]
\,\,\,\,\,\,.
\eea
Next, the Lagrangian transforms as 
\bea
L_{\mathrm{fermion}}   \longrightarrow L_{\mathrm{fermion}}   
+ i \sqrt{-g}\,\bar\psi E_c^\mu  \left[ -\frac{3}{4}\,\gamma^c\,\dd_\mu\sigma
+ \frac{1}{4}\left(\eta^{bc}\,\gamma^d \Sigma_{db} - \eta^{ac}\,\gamma^d \,\Sigma_{ad}\right)\dd_\mu\sigma\right]\psi
 \,\,\,\,.
\label{scale-fermion}
\eea 
Using the properties  of the Dirac matrices as in (\ref{Dirac}) and recalling that $\gamma_a =\eta_{ab}\,\gamma^b$, one finds that
\bea
\eta^{bc}\,\gamma^d \Sigma_{db} - \eta^{ac}\,\gamma^d \,\Sigma_{ad} = 3\,\gamma^c \,\,\,\,\,\,.
\eea
This means that the terms involving $\dd_\mu\sigma$ in (\ref{scale-fermion}) cancel and we have 
\bea
L_{\mathrm{fermion}}   \longrightarrow L_{\mathrm{fermion}}  \,\,\,\,\,\,.
\eea
Finally, a typical Yukawa interaction in the Standard Model is of the form
\be
L_{\mathrm{Yukawa}} =  \sqrt{-g}\left(\bar l\,H^\dagger\,L + \bar L\,H\,l\right)\,\,\,\,\,\,,
\ee
where $L$ is a fermionic doublet and $l$ is a fermionic singlet. It is clear that this 
interaction term is invariant under 
$g_{\mu\nu}  \longrightarrow   e^{\sigma}\, g_{\mu\nu}$,  $L\longrightarrow    e^{-\frac{3}{4}\sigma}\,L$, 
$l\longrightarrow    e^{-\frac{3}{4}\sigma}\,l$ and 
$H \longrightarrow   e^{-\frac{1}{2}\sigma}\, H$.


\begin{thebibliography}{99}

\bibitem{rev1}
E. Scholz, {\it The Unexpected Resurgence of Weyl Geometry in late 20th-Century Physics}, Einstein Stud. {\bf 14} (2018) 261-360, arXiv:1703.03187 [math.HO]. 


\bibitem{rev2}
N. Rosen, {\it Weyl’s geometry and physics}, Foundations of
Physics {\bf 12}, 213 (1982).


\bibitem{rev3}
 J. T. Wheeler, {\it Weyl geometry}, Gen. Rel. Grav. {\bf 50} (2018) 80,  arXiv:1801.03178 [gr-qc].


\bibitem{rev4}
Alfredo Iorio,  L. O'Raifeartaigh and I. Sachs,
{\it Weyl gauging and conformal invariance}, Nucl. Phys. {\bf B 495} (1997) 433-450, hep-th/9607110 [hep-th].

\bibitem{rev5}
Lochlainn O'Raifeartaigh and Norbert Straumann, {\it Gauge theory: Historical origins and some modern developments}, 
Rev. Mod. Phys. {\bf 72} (2000) 1-23.


\bibitem{ob1}
M. Israelit, {\it A Weyl-Dirac cosmological model with DM and DE}, Gen. Rel. Grav. {\bf 43} (2011) 751-775, arXiv:1008.0767 [gr-qc].
  	
  	
\bibitem{ob2}
 D. M. Ghilencea, {\it Gauging scale symmetry and inflation: Weyl versus Palatini gravity}, 
Eur. Phys. J. {\it C 81}, 510 (2021), arXiv:2007.14733 [hep-th].

\bibitem{ob3}
D. M. Ghilencea, {\it Non-metric geometry as the origin of mass in gauge theories of scale invariance}, 
Eur. Phys. J. {\bf C 83}, 176 (2023), arXiv:2203.05381 [hep-th].

\bibitem{ob4}
P. Burikham, T. Harko, K. Pimsamarn, and S. Shahidi, {\it Dark matter as a Weyl geometric effect}, 
Phy.  Rev. {\bf D 107}, 064008 (2023), arXiv:2302.08289 [gr-qc].

\bibitem{ob5}
Marius A. Oancea1  and Tiberiu Harko, {\it Weyl geometric effects on the propagation of light in gravitational fields}, 
(2023), arXiv:2305.01313 [gr-qc].

\bibitem{ob6}
Maria Cr\v{a}ciuna and  Tiberiu Harkob, {\it Testing Weyl geometric gravity with the SPARC galactic rotation curves database}, 
Phys. Dark Univ. {\bf 43} (2024) 101423,  arXiv:2311.16893 [gr-qc].

\bibitem{ob7}
P. D. Mannheim and J. G. O’Brien, {\it Fitting galactic rotation curves with conformal gravity and a global quadratic potential}, 
Phys. Rev. {\bf D 85} (2012) 124020,  arXiv:1011.3495 [astro-ph.CO] 	

\bibitem{ob8}
J. G. O’Brien and  P. D. Mannheim, {\it Fitting dwarf galaxy rotation curves with conformal gravity}, 
Mon. Not. R. Astron. Soc. {\bf 421} (2) (2012) 1273–1282, arXiv:1107.5229 [astro-ph.CO].

\bibitem{ob9}
Cemsinan Deliduman, Oguzhan Kasikci and Baris Yapiskan, 
{\it Flat galactic rotation curves from geometry in Weyl gravity}, Astrophys. Space Sci. 365 (3) (2020) 51, arXiv:1511.07731 [gr-qc].

\bibitem{ob10}
M. Hobson and A. Lasenby, {\it Conformally-rescaled Schwarzschild metrics do not predict flat galaxy rotation curves}, 
Eur. Phys. J. {\bf C 82} (7) (2022) 585, arXiv:2206.08097 [gr-qc].

\bibitem{ob11}
Z. Haghani and  T. Harko, {\it Compact stellar structures in Weyl geometric gravity}, Phys. Rev. {\bf D 107} (2023) 064068, arXiv:2303.10339 [gr-qc].

\bibitem{ob12}
 M. A. Oancea, T. Harko,{\it  Weyl geometric effects on the propagation of light in gravitational fields}, (2023), arXiv:2305.01313 [gr-qc].

\bibitem{ob13}
P. K. Aluri, P. Jain and N. K. Singh, {\it Dark Energy and Dark Matter in General Relativity with local scale invariance}, 
Mod. Phys. Lett.{\bf  A 24} (2009) 1583-1595, arXiv:0810.4421 [hep-ph]. 

\bibitem{ob14}
P. K. Aluri, P. Jain, S. Mitra, S. Panda and N. K. Singh, {\it Constraints on the Cosmological Constant due to Scale Invariance}, 
Mod. Phys. Lett. {\bf A 25} (2010) 1349-1364, arXiv:0909.1070 [hep-ph].

\bibitem{ob15}
D. M. Ghilencea and T. Harko, {\it Cosmological evolution in Weyl conformal geometry}, (2021), arXiv:2110.07056 [gr-qc].


\bibitem{ob16}
Y. Tang and Y. L. Wu, {\it Weyl Symmetry Inspired Inflation and Dark Matter}, Phys. Lett. {\bf B 803} (2020) 135320,  arXiv:1904.04493 [hep-ph].

\bibitem{ob17}
R. Kallosh and A. Linde, {\it Universality Class in Conformal Inflation}, JCAP {\bf 07} (2013) 002, arXiv:1306.5220 [hep-th].

\bibitem{ob18}
R. Jackiw and S. Y. Pi, {\it Fake Conformal Symmetry in Conformal Cosmological Models}, Phys. Rev. {\it D 91} (2015) 067501,  arXiv:1407.8545 [gr-qc].

\bibitem{ob19}
R. Jackiw and S. Y. Pi, {\it New Setting for Spontaneous Gauge Symmetry Breaking?}, Fundam. Theor. Phys. {\bf 183} (2016) 159, arXiv:1511.00994 [hep-th].

\bibitem{ob20}
A. Paliathanasis, G. Leon and J. D. Barrow, {\it Inhomogeneous spacetimes in Weyl integrable geometry with matter source},” 
Eur. Phys. J. {\bf C 80} (2020) 731, arXiv:2006.01793 [gr-qc].

\bibitem{ob21}
J. Miritzis, {\it Acceleration in Weyl integrable spacetime},  Int. J. Mod. Phys. {\bf D 22} (2013) 1350019, arXiv:1301.5696 [gr-qc].

\bibitem{ob22}
E. Scholz, {\it MOND-like acceleration in integrable Weyl geometric gravity},  Found. Phys. {\bf 46} (2016) 176-208, arXiv:1412.0430 [gr-qc].

\bibitem{ob23}
R. Aguila, J. E. Madriz Aguilar, C. Moreno and M. Bellini, {\it Present accelerated expansion of the universe from new Weyl-Integrable gravity approach}, 
Eur. Phys. J. {\bf C 74} (2014) 3158,  arXiv:1408.4839 [gr-qc]. 

\bibitem{ob24}
P. G. Ferreira, C. T. Hill and G. G. Ross, {\it Weyl Current, Scale-Invariant Inflation and Planck Scale Generation}, 
Phys. Rev. {\bf D 95} (2017) 043507, arXiv:1610.09243 [hep-th].

\bibitem{ob25}
P. G. Ferreira, C. T. Hill and G. G. Ross, {\it Inertial Spontaneous Symmetry Breaking and Quantum Scale Invariance}, 
Phys. Rev. {\bf D 98} (2018) 116012, arXiv:1801.07676 [hep-th].

\bibitem{ob26}
P. G. Ferreira, C. T. Hill and G. G. Ross, {\it No fifth force in a scale invariant universe}, Phys. Rev. {\bf D 95} (2017) 064038, arXiv:1612.03157 [gr-qc]. 

\bibitem{ob27}
P. G. Ferreira, C. T. Hill and G. G. Ross, {\it Scale-Independent Inflation and Hierarchy Generation}, 
Phys. Lett. {\bf B 763} (2016) 174-178, arXiv:1603.05983 [hep-th].



\bibitem{sm1}
D. M. Ghilencea, {\it Spontaneous breaking of Weyl quadratic gravity to Einstein action and Higgs potential}, JHEP {\it 03} (2019) 049, arXiv:1812.08613 [hep-th].

\bibitem{sm2}
D. M. Ghilencea and H. M. Lee, {\it Weyl gauge symmetry and its spontaneous breaking in the standard model and inflation}, Phys. Rev. {\bf D 99}, 115007 (2019), arXiv:1809.09174 [hep-th].

\bibitem{sm3} 
D. M. Ghilencea, {\it Stueckelberg breaking of Weyl conformal geometry and applications to gravity}, Phys. Rev. {\bf D 101}, 045010 (2020), arXiv:1904.06596 [hep-th].

\bibitem{sm4} 
D. M. Ghilencea and C. T. Hill, {\it Standard Model in conformal geometry: local vs gauged scale invariance}, Annals Phys. {\bf 460} (2024) 169562, arXiv:2303.02515 [hep-th].


\bibitem{sm5} 
I. Bars, P. J. Steinhardt and N. Turok, {\it Cyclic cosmology, conformal symmetry and the metastability of the Higgs}, Phys.  Let. {\bf  B 726}, 50-55 (2013), arXiv:1307.8106 [gr-qc].

\bibitem{sm6}  
D. M. Ghilencea, {\it Standard model in Weyl conformal geometry}, Eur. Phys. J. {\bf C 82} (2022) 23,  	arXiv:2104.15118 [hep-ph].

\bibitem{sm7}
M. de Cesare, J. W. Moffat and M. Sakellariadou, {\it Local conformal symmetry in non-Riemannian geometry and the origin of physical scales},” Eur. Phys. J. {\bf C 77} (2017) 605, 
arXiv:1612.08066 [hep-th].

\bibitem{sm8}
H. Nishino and S. Rajpoot, {\it Implication of Compensator Field and Local Scale Invariance in the Standard Model}, 
Phys. Rev. {\bf D 79} (2009) 125025,  arXiv:0906.4778 [hep-th].

\bibitem{sm9}
E. I. Guendelman, H. Nishino and S. Rajpoot, {\it Local scale-invariance breaking in the standard model by two-measure theory}, 
Phys. Rev. {\bf D 98} (2018) 055022.

\bibitem{sm10}
H. C. Ohanian, {\it Weyl gauge-vector and complex dilaton scalar for conformal symmetry and its breaking}, 
Gen. Rel. Grav. {\bf 48} (2016) 25, arXiv:1502.00020 [gr-qc].

\bibitem{sm11}
I. Quiros, {\it On the physical consequences of a Weyl invariant theory of gravity}, (2020), arXiv:1401.2643 [gr-qc].

\bibitem{sm12}
P. Jain, S. Mitra and N. K. Singh, {\it Cosmological Implications of a Scale Invariant Standard Model}, 
JCAP {\bf 03} (2008) 011,  arXiv:0801.2041 [astro-ph].

\bibitem{sm13}
J. Garcia-Bellido, J. Rubio, M. Shaposhnikov and D. Zenhausern, {\it Higgs-Dilaton Cosmology: From the Early to the Late Universe}, Phys. Rev. {\bf D 84} (2011) 123504, 
arXiv:1107.2163 [hep-ph]. 




\bibitem{c1}
J.-Z. Yang, S. Shahidi, and T. Harko, {\it Black hole solutions in the quadratic Weyl conformal geometric theory of gravity}, 
Eur. Phys. J. {\bf C 82} (2022) 1171, 	arXiv:2212.05542 [gr-qc] 

\bibitem{c2}
P. D. Mannheim and D. Kazanas, {\it Exact Vacuum Solution to Conformal Weyl Gravity and Galactic Rotation Curves}, 
Astrophys. J. {\bf 342} (1989) 635-638. 

\bibitem{c3}
M.  Hohmann, C.  Pfeifer, M.  Raidal and  H. Veerm\"aäe, {\it Wormholes in conformal gravity }, JCAP {\bf 10} (2018) 003,  arXiv:1802.02184 [gr-qc].


\bibitem{ex1}
Boris Kors and Pran Nath,
{\it A Stueckelberg Extension of the Standard Model}, Phys. Lett. {\bf B586} (2004) 366-372,  arXiv:hep-ph/0402047.

\bibitem{ex2}
Daniel Feldman, Zuowei Liu  and Pran Nath, {\it Stueckelberg $Z'$ extension with kinetic mixing and millicharged dark matter from the hidden sector}, 
Phys. Rev. {\bf D75} (2007) 115001, arXiv:hep-ph/0702123.




\bibitem{kaiser}
David Kaiser, {\it Primordial Spectral Indices from Generalized Einstein Theories}, Phys. Rev. {\bf D 52} (1995) 4295-4306,	arXiv:astro-ph/9408044.


\bibitem{bezrukov1}
Fedor L. Bezrukov and Mikhail Shaposhnikov, {\it The Standard Model Higgs boson as the inflaton}, Phys. Lett. {\bf B 659} (2008) 703,  arXiv:0710.3755 [hep-th].

\bibitem{bezrukov2}

Fedor Bezrukov, {\it The Higgs field as an inflaton}, Class. Quantum Grav. {\bf 30} (2013) 214001, arXiv:1307.0708 [hep-ph].


\bibitem{me}
N.  Mohammedi, {\it On Higgs inflation in non-minimally coupled models of gravity},  Phys. Lett. {\bf B 831} (2022) 137180, arXiv:2202.05696 [hep-th].


\end{thebibliography}
\end{document}